\documentclass[aps,prl,twocolumn,showpacs,superscriptaddress]{revtex4}
\usepackage{amsfonts}

\usepackage{graphicx}
\usepackage{subfigure}
\usepackage{braket}
\usepackage{amsmath}
\usepackage{bm}

\begin{document}

\title{Vibrational interference of Raman and high-harmonic generation pathways}
\author{Zachary~B.~Walters}
\email{zwalters@gmail.com}
\affiliation{Department of Physics and JILA, University of
Colorado, Boulder, Colorado 80309-0440, USA}
\author{Stefano Tonzani}
\email{tonzani@gmail.com}
\affiliation{Nature Publishing Group, 4 Crinan Street, London N1 9XW, UK}
\author{Chris~H.~Greene}
\email{chris.greene@colorado.edu}
\affiliation{Department of Physics and JILA, University of
Colorado, Boulder, Colorado 80309-0440, USA}
\date{\today}

\begin{abstract}
Experiments have shown that the internal vibrational state of a
molecule can affect the intensity of high harmonic light generated
from that molecule.  This paper presents a model which explains this
modulation in terms of interference between different vibrational
states occurring during the high harmonic process.  In addition, a
semiclassical model of the continuum electron propagation is developed
which connects with rigorous treatments of the electron-ion scattering.
\end{abstract}

\maketitle

\section{Introduction}

In the usual three step model of high harmonic generation,
HHG is treated as a purely electronic process.  A single
active electron tunnels free of a parent molecule and eventually
scatters from it, but the molecule itself is treated in essentially
the same way as a lone atom; no more than a complicated potential
influencing the active electron.  However, molecules differ from atoms
in an essential way because they possess non electronic internal
degrees of freedom, which can themselves be affected by the high
harmonic process.

The possibility that such internal degrees of freedom could play a
detectable role in HHG is intriguing, because the intrinsic timescale
in HHG -- the time necessary to ionize, propagate and rescatter -- is
only half a laser cycle.  This is faster than many chemically
interesting processes, holding out the possibility that HHG could
serve as a probe of molecular motion.  Alternatively, tailoring the
state of a molecule prior to HHG could serve to give additional
control over the generated light.

These issues were brought to the fore by an experiment at JILA
\cite{wagner2006}.  In the experiment, a high harmonic generating
laser pulse was preceded by a weaker pulse whose effect was to
stimulate Raman-active vibrations in SF$_{6}$ molecules.  Varying the
delay between the two pulses was observed to modulate the
intensity of the HHG light generated by the second pulse.  Moreover,
the modulation corresponds to the frequencies of the
Raman-active normal modes stimulated by the first pulse.  None of the
3 non-Raman-active vibrational modes of SF$_{6}$ were detected in the
modulated signal.

This paper presents a quantum mechanical model of high
harmonic generation in molecules.  This model provides a framework to
interpret the observed modulation of high harmonic intensities
observed in the JILA experiment, and is easily extended to systems
with more complicated dynamics.  Secondly, it presents a version of
the three step model which has been improved for the purpose of
treating HHG in molecules with relevant internal degrees of freedom.
Finally, the modulations predicted by this improved model are compared
with the modulations observed in the JILA experiment.  This paper
recapitulates and extends work which originally appeared in
\cite{walters2007a}. 

\section{The Vibrational Wavefunction of the Molecule}

An important difference between atomic and molecular systems, probed
by the aforementioned experiment\cite{wagner2006} is the presence of
vibrational degrees of freedom in the latter.
For an $M$-atom molecule, this corresponds to $N=3M-6$
($N=3M-5$ for linear molecules) internal degrees of freedom, which can
be expressed in normal mode coordinates.  The vibrational wavefunction
of the molecule can then be expanded as the product of simple harmonic
oscillator basis functions in each of the normal modes of the molecule:
\begin{equation}
\ket{\psi_{\text{vib}}}=\sum_{n_{1},n_{2},\ldots n_{N}=0}^{\infty}
A_{n_{1},n_{2},\ldots,n_{N}}\ket{n_{1},n_{2},\ldots n_{N}}
\end{equation}
where
\begin{equation}
\bf{a^{(i)\dag}}\bf{a^{(i)}}\ket{n_{1} \ldots n_{N}}=n_{i} \ket{n_{1}\ldots n_{N}}
\end{equation}
where $\ket{n_{1} n_{2} \ldots n_{N}}$ is the outer product of simple
harmonic oscillator state $\ket{n_{1}}$ in the first normal mode,
$\ket{n_{2}}$ in the second normal mode, and so on.

For the purposes of this paper, all operators will be expanded to
first order in raising $\bf{a^{\dag(i)}}$ and lowering $\bf{a^{(i)}}$ operators
in each normal mode $i$. At this level of approximation, the evolution of the
vibrational wavefunction becomes separable, and the overall
coefficient $A_{n_{1} n_{2}\ldots n_{N}}(t)$  can be factored into the
product of individual coefficients $a_{n_{i}}(t)$:
\begin{equation}
A_{n_{1} n_{2}\ldots n_{N}}(t)=a_{n_{1}}(t) a_{n_{2}}(t) \ldots
a_{n_{N}}(t). 
\label{eq:separability}
\end{equation}
(Note that the coefficients $a_{n_{i}}(t)$ should not be confused with lowering
operators $\bf{a^{(i)}}$.)

In the interests of simplicity and clarity, the remainder of this
paper will use a 1-D picture, describing the evolution of the
vibrational wavefunction for each single normal mode.  The concepts from
the 1-D model extend readily to the higher intensity regime of coupled
modes, in the event
that operators involving two or more raising or lowering operators
become important. \footnote{The lack of modulation at combination frequencies
like $2 \omega_{1}$, $\omega_{1}+\omega_{2}$,$\omega_{1}-\omega_{2}$,
etc. suggests but does not prove that such higher-order terms are not
important in discussing the JILA experiment.}

\section{Vibrational Interference}

\begin{figure}
\begin{center}
\includegraphics[width=3.375in]{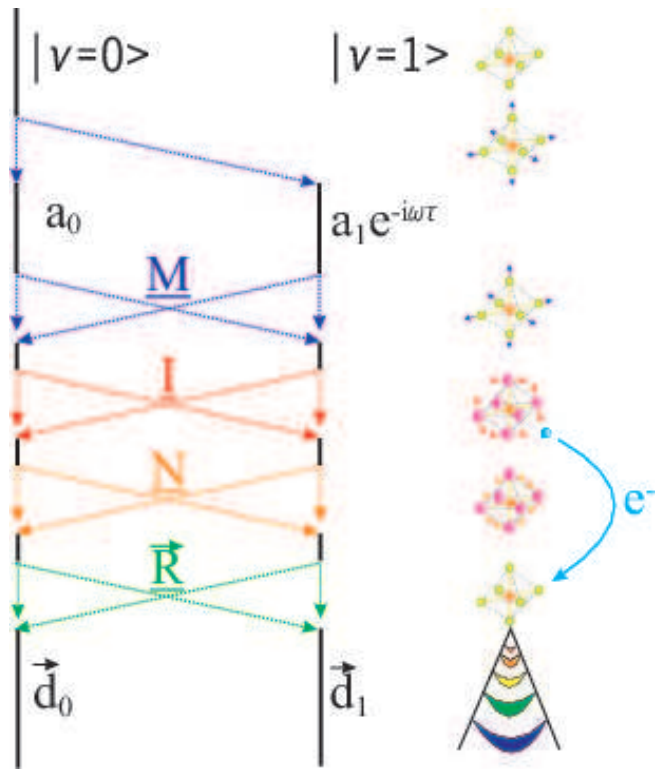}
\end{center}
\caption{The vibrational interference model \cite{walters2007a} in one dimension.  The
  molecule ends the first (Raman) pulse in a superposition of the
  $\nu=0$ and $\nu=1$ vibrational states.  After a time delay, the two
  vibrational states are mixed by stimulated Raman scattering
  (transfer matrix \underline{M}), ``hopping'' during ionization
  (\underline{I}) and recombination (\underline{R}), as well as
  evolution of the ionic wavefunction while the electron is away
  (\underline{N}).  Interference between adjacent vibrational
  states modulates the high harmonic signal.}
\vskip 0.5in
\label{fig:crossings}
\end{figure}

Although high harmonic generation is primarily an electronic process,
the vibrational state of the molecule can affect the harmonic
intensity.  This occurs because at several points in the high harmonic
process, the 
vibrational wavefunction has amplitudes either to stay unchanged or to
``hop'' up or down from simple harmonic oscillator state $\ket{n}$ to
states $\ket{n \pm 1}$, much in the same way that a photon in a
beamsplitter has amplitudes to take two or more paths.   As in
a beamsplitter experiment, two or more indistinguishable pathways
interfere with one another and modulate the output signal detectably. 
The multiple pathways at play
in the high harmonic process are diagrammed in Figure
\ref{fig:crossings}. 

\paragraph{Raman Excitation}
The first opportunity to change vibrational states occurs when the
vibrationally cold molecules are subjected to the weak, off-resonant
initial pulse.  This causes the molecule to undergo
stimulated Raman scattering.  At the start of the first pulse the
molecule is
in the $\ket{0}$ vibrational state, and it is then driven into a coherent
superposition of the zeroth and first vibrational states 
\begin{equation}
\ket{\psi_{\text{vib}}}=a_{0}(0)\ket{0}+a_{1}(0)\ket{1}
\end{equation}
by the end of the pulse.  (For the pulse length and intensity used in
the JILA experiment, calculations show no appreciable population of the
$\ket{2}$ or higher states after the Raman pulse.  Accordingly, the
$\ket{2}$ and higher states have been dropped from this analysis.)
Because only one normal mode is used in this analysis, the $N$ quantum
number has been dropped, so that $a_{0}=a_{1_{0}}(0)$ and
$a_{1}=a_{1_{1}}(0)$.  $t=0$ is chosen at some time after the end of
the first pulse, when the stimulated Raman scattering is over.

The vibrational state coefficients follow equations of motion given by
\begin{equation}
\begin{split}
i\dot{a}_{n_{i}}(t) =\omega _{i}\left(n_{i}+\frac{1}{2}\right)a_{n_{i}}(t)
-\frac{1}{2}\sum_{A,B}E_{A}(t)E_{B}(t) \times  \label{eq:stimulatedraman} \\
\left[ \alpha _{AB}a_{n_{i}}+\partial
_{i}\alpha _{AB}(\sqrt{n_{i}+1}a_{n_{i}+1}+\sqrt{n_{i}}a_{n_{i}-1})\right] .
\end{split}
\end{equation}
Here $\omega _{i}$ is the normal mode frequency, indices $A$ and $B$
run over $\{x,y,z\}$, $E_{A}(t)$ is the component of the electric field in
the (body-frame) $A$ direction at time $t$, $Q_{i}$ is the normalized
displacement associated with normal mode $i$ and $\alpha
_{AB}(Q_{1},Q_{2},...)$ is the polarizability tensor of the molecule. These
equations of motion have off-diagonal elements only if $\partial _{i}\alpha
_{AB}\equiv \left( 2m\omega _{i}\right) ^{-1/2}\partial \alpha
_{AB}/\partial Q_{i}|_{Q_{i}=0}\neq 0$, which is the condition for a mode to be
Raman active. The polarizability tensor and its derivatives
 are found by performing an unrestricted Hartree-Fock calculation \cite{g98}
using the aug-cc-pVTZ basis set.
\cite{dunning1989}

Between the two pulses, the $\ket{0}$ and $\ket{1}$ states evolve as
eigenstates of the simple harmonic oscillator Hamiltonian.  The states
are then mixed once again by stimulated Raman scattering during the
high harmonic generating pulse.  These effects are approximated by a
unitary $2\times2$  transfer matrix $\underline{M}$, where
$M_{ij}$ is the amplitude to be in state $\ket{i}$ at the instant of
ionization after beginning the second pulse in state $\ket{j}$.

\paragraph{Ionization and Recombination}
The vibrational wavefunction evolves further during each of
the three steps --ionization, propagation, and recombination--of the
three step model.  The molecule's vibrational state hops up or
down a level during ionization, evolves while the propagating electron
is away from the molecule, and hops once again when the electron
recombines with the parent ion.

These hopping amplitudes arise because ionization and recombination,
commonly thought of as purely electronic processes, are both strongly
modulated by molecular distortions.  This is the simplest way in which
the internal degrees of freedom in molecules allows for behavior that
has no analogue in atomic systems.  Nonzero derivatives of ionization
and recombination amplitudes translate directly into amplitudes for
the molecule to change its vibrational state during these processes.
Taylor-expanding the ionization operator about the equilibrium
configuration of the neutral molecule,
\begin{equation}
\hat{I}=\hat{I}|_{\text{eq}}+\frac{\partial \hat{I}}{\partial Q}Q +\mathbb{O}(Q^{2}),
\end{equation}
using the identity $Q=(\bf{a}+\bf{a^{\dag}})/\sqrt{2 m \omega}$ and substituting
$I_{0}=\hat{i}|_{\text{eq}}$, $I_{1}=(2 m \omega)^{-1/2}
\frac{\partial \hat{I}}{\partial Q}$, the ionization operator can be
rewritten
\begin{equation}
\hat{I}=I_{0}+I_{1}(\bf{a}+\bf{a^{\dag}}).
\end{equation}
Identical logic gives the recombination dipole vector
operator
\begin{equation}
\hat{\vec{R}}=\vec{R}_{0}+\vec{R}_{1}(\bf{a}+\bf{a^{\dag}})
\end{equation}

\paragraph{Vibrational Dynamics of the Parent Ion}
Between the times of ionization and recombination, the evolution of
the internal state of SF$_{6}^{+}$ is quite complicated.  This is
because SF$_{6}$ has three degenerate orbitals at the point of maximum
symmetry.  Thus, at any nuclear configuration {\em near} this maximum
symmetry point, these three SF$_{6}^{+}$ orbitals are very
nearly degenerate, and are mixed with one another strongly
by molecular distortions.  Orbital degeneracies can be broken and
orbital energies can cross even with relatively small distortions.  Because
of this, it is necessary to treat this interplay between electronic
and vibrational states when describing the dynamics of SF$_{6}^{+}$ in
the vicinity of the maximum symmetry point.

In its maximum symmetry configuration, SF$_{6}^{+}$ belongs to the
O$_{h}$ point group, with three degenerate T$_{1g}$ orbitals which
transform like axial vectors $\hat{x}$, $\hat{y}$ and $\hat{z}$.
When the molecule is distorted away from the maximum symmetry point
via either an E$_{g}$ or a T$_{2g}$ distortion, the triple degeneracy
breaks up into three nondegenerate electronic orbitals.  The fully
symmetric A$_{1g}$ or ``breathing'' mode preserves the triple
degeneracy.

The full vibronic (vibrational-electronic) Jahn-Teller coupling matrix
for a triply degenerate system is given by\cite{estreicher,moffitt,bersuker}
\begin{equation}
H_{T}=
\begin{pmatrix}
g_{1}-g_{\theta}+\sqrt{3}g_{\epsilon} & g_{\zeta} & g_{\eta} \\
g_{\zeta} & g_{1}-g_{\theta}-\sqrt{3} g_{\epsilon} & g_{\xi} \\
g_{\eta} & g_{\xi} & g_{1}+2 g_{\theta}
\end{pmatrix}
\label{eq:vibroniccouplingmatrix}
\end{equation}  

This matrix represents the coupling between the states with $\hat{x}$,
$\hat{y}$, and $\hat{z}$ symmetry, caused by vibrational operators,
so that $g_{\zeta}$ represents the off-diagonal
coupling between the states with $\hat{x}$ and
$\hat{y}$ symmetry, $g_{\eta}$ the off-diagonal coupling between the states
with $\hat{x}$ and $\hat{z}$ symmetry, and $g_{\xi}$ represents the
off-diagonal coupling between the states with $\hat{y}$
and $\hat{z}$ symmetry. 

Defining the E$_{g}$ normal mode coordinates  $Q_{\theta}$ and
$Q_{\epsilon}$, which transform like $2 z^{2}-x^{2}-y^{2}$ and
$x^{2}-y^{2}$ respectively, and the T$_{2g}$ normal mode coordinates
$Q_{\xi}$, $Q_{\eta}$ and $Q_{\zeta}$ respectively as the coordinates that
transform like $yz$,$xz$, and $xy$, the functions $g_{i}$ are given by
\cite{estreicher} 
\begin{eqnarray}
g_{1}=&\frac{1}{2} K_{E} (Q_{\theta}^{2}+Q_{\epsilon}^{2}) +
\frac{1}{2} K_{T}(Q_{\xi}^{2}+Q_{\eta}^{2}+Q_{\zeta}^{2}) \\
g_{\theta}=&\frac{1}{\sqrt{3}} V_{E_{g}} Q_{\theta} +
N_{E}(Q_{\epsilon}^{2}-Q_{\theta}^{2})+N_{1}(2
Q_{\zeta}^{2}-Q_{\xi}^{2}-Q_{\eta}^{2}) \\
g_{\epsilon}=&\frac{1}{\sqrt{3}} V_{E_{g}}Q_{\epsilon} + 2 N_{E}
Q_{\theta} Q_{\epsilon}+ \sqrt{3}N_{1}(Q_{\xi}^2-Q_{\eta}^{2}) \\
g_{\xi}=&V_{T_{2g}} Q_{\xi} +N_{T} Q_{\eta}Q_{\zeta}+N_{2}
Q_{\xi}(\sqrt{3}Q_{\epsilon}-Q_{\theta}) \\
g_{\eta}=&V_{T_{2g}}Q_{\eta} + N_{T} Q_{\zeta} Q_{\xi} +
N_{2}Q_{\eta}(-\sqrt{3}Q_{\epsilon}-Q_{\theta}) \\
g_{\zeta}=&V_{T_{2g}} Q_{\zeta} +N_{T} Q_{\xi}Q_{\eta}+2 N_{2} Q_{\zeta}Q_{\theta}
\end{eqnarray}

\begin{figure}
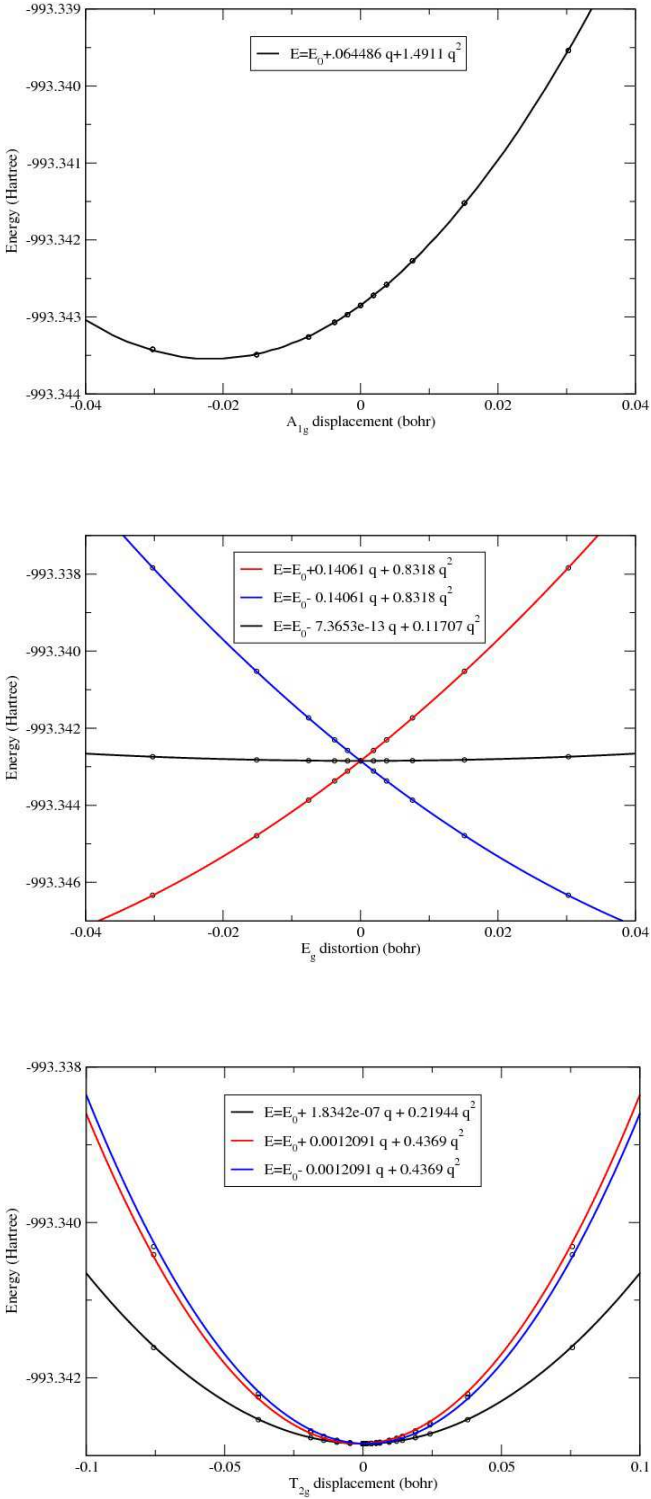

\begin{center}
\includegraphics[width=3.375in]{A1g_jtenergies_distancebohr.eps2}
\vskip 0.5in
\includegraphics[width=3.375in]{Eg_jtenergies_distancebohr.eps2}
\vskip 0.5in
\includegraphics[width=3.375in]{T2g_jtenergies_distancebohr.eps2}
\end{center}
\caption{a)Spherically symmetric $A_{1g}$ (breathing
  mode) distortions change electronic state energies, but preserve the
  triple degeneracy.  Non symmetric E$_{g}$ (b) and T$_{2g}$ (c)
  distortions break the triple degeneracy of SF$_{6}^{+}$ at the
  maximum symmetry point.  Adiabatic energies are fitted to the
  eigenvalues of the vibronic coupling matrix (equation
  \ref{eq:vibroniccouplingmatrix}) to solve for the vibronic coupling
  constants.}
\label{fig:adiabaticenergiesvsdistortion}
\end{figure}

These constants were found by performing a CASSCF state-averaged
calculation for the three lowest energy states of SF$_{6}^{+}$ for
various displacements of the molecule away from the maximum symmetry
configuration.  These energies are shown in Figure
\ref{fig:adiabaticenergiesvsdistortion}.  The CASSCF calculations were
carried using a basis of 
Hartree-Fock orbitals calculated for neutral SF$_{6}$.  These three
adiabatic energies were then fitted to the eigenvalues of the diabatic
Jahn-Teller coupling matrix.  This process yielded $V_{T_{2g}}$=.001209 H/bohr,
$V_{E_{g}}$=.1406 H/bohr,
$N_{1}$=-.0362 H/bohr$^2$,
$K_{T_{2g}}$=.7288 H/bohr$^2$,
$K_{E_{g}}$=1.8486 H/bohr$^2$.  For the A$_{1g}$ mode, which does not enter
into the vibronic Hamiltonian,  an adiabatic potential
$E=V_{A_{1g}}Q_{A_{1g}}+\frac{1}{2} K_{A_{1g}}Q_{A_{1g}}^2$, with
$V_{A_{1g}}$=.0645 H/bohr, 
$K_{A_{1g}}$=2.98 H/bohr$^2$ gives the potential energy surface for all
three electronic states.

Here there is a significant distinction between distortions of type
$E_{g}$, which break the triple degeneracy but have no linear
off-diagonal terms, and distortions of type T$_{2g}$, which do
contribute to off-diagonal coupling.  For small distortions, an
adiabatic electronic state of an E$_{g}$ distorted molecule will have
the same symmetry -- $\hat{x}$, $\hat{y}$, $\hat{z}$ -- as the diabatic
electronic states.  The adiabatic electronic states of a
T$_{2g}$-distorted molecule, on the other hand, are linear
combinations of the diabatic orbitals.
An important simplification is that $V_{T_{2g}}$, controlling off-diagonal
coupling between different electronic states, is small in SF$_{6}^{+}$
and can be neglected for the short time between ionization and recombination.

The evolution of the ionic wavefunction is calculated in the
vibrational basis of the neutral molecule, but using potential energy
surfaces calculated for the ion.  Potential energy curves are found to
quadratic order in $Q$ using quantum chemistry calculations, then
expressed in terms of raising and lowering operators by
substituting $Q=(2 m \omega)^{-1/2}(\bf{a}+\bf{a^{\dag}})$, $Q^{2}=(2 m
\omega)^{-1}(\bf{a}+\bf{a^{\dag}})(\bf{a}+\bf{a^{\dag}})$.  All terms up to linear
in raising and lowering operators are then used to integrate the time-dependent
Schr\"odinger equation to find a transfer matrix $\underline{N}$
describing the evolution of the ionic wavefunction between ionization
and recombination.

\paragraph{Modulation of Harmonic Intensity}
In the two-state model used here, the $i-$th vibrational wavefunction of
the neutral molecule after recombination has occurred is 
$\ket{\psi_{\text{vib}}}=d_{0}\ket{0}+d_{1}\ket{1}$, where 
\begin{equation}
\begin{pmatrix}
\vec{d}_{0} & \vec{d}_{1}
\end{pmatrix}
=
\begin{pmatrix}
a_{0}(0) & a_{1}(0) e^{-i \omega \tau}
\end{pmatrix}
\underline{M}^{T}\underline{I}^{T}\underline{N}^{T}\underline{\vec{R}}^{T}.
\end{equation}
Here, e.g. $\underline{M}^{T}$ denotes the transpose of matrix $\underline{M}$.

The number of photons
emitted in a given harmonic is proportional to
$\vec{d}_{0}\cdot\vec{d}^{*}_{0}+\vec{d}_{1}\cdot{\vec{d}^{*}_{1}}$. 
The high harmonic intensity is a sum over all Raman
active modes $i$: 
\begin{equation}
P(\tau )=P_{0}+\Sigma _{i}P_{1}^{(i)}\cos \left(\omega
_{i}\tau+\delta _{i}\right)
\label{eq:intensity}
\end{equation}.  
The static $P_{0}$ primarily results from terms of the form
$a_{0}(0)^{*}a_{0}(0)$, while $P_{1}$ results from terms of the form
$a_{0}(0)a_{1}(0)^{*}e^{i \omega \tau}$ and $a_{1}(0)a_{0}(0)^{*}e^{-i \omega
  \tau}$.  Defining
$\underline{W}=\underline{M}^{\dag}\underline{I}^{\dag}
\underline{N}^{\dag} 
\underline{\vec{R}}^{\dag}\cdot \underline{\vec{R}NIM}$,
$P_{0}=a_{0}(0)^{*}W_{00}a_{0}(0)$ and $P_{1}\cos \left( \omega
t+\delta \right)=\frac{1}{2}(a_{1}^{*}e^{i
  \omega \tau}W_{10}a_{0}(0)+\text{c.c.})$.  Since $I_{1}$ and
$R_{1}$ are small relative to $I_{0}$ and $R_{0}$, only their
first-order terms are kept.

\section{Describing the Continuum Electron}


The evolution of the continuum electron wavefunction is strongly
influenced by interactions with both the parent ion and the driving laser.
When the
electron first tunnels free out of the parent molecule, its wavefunction
is determined by both the molecular potential and the electric field
of the laser.  Once free, it propagates in the time-varying field of
the laser while feeling a weak force due to Coulomb attraction to the
parent ion.  Finally it recollides with the parent ion, and is once
again strongly distorted by the molecular potential.


A full solution of the time dependent Schr\"odinger equation for this
process would be computationally demanding for complicated molecules
such as SF$_{6}$.  In addition, much of the information in the
continuum wavefunction is not relevant to the HHG problem: only a
small part of the wavefunction overlaps with the unoccupied orbital
into which the rescattering electron recombines.

One frequently used treatment which avoids the
complications of the full time-dependent Schr\"odinger equation
\cite{corkum93,lewenstein94} is based on a classical or semiclassical
propagation of the continuum electron, ignoring the ionic Coulomb
potential.  The returning electron wavefunction is the approximated as
a plane wave throughout the recombination process.  This approach has
been successful in describing the high harmonic cutoff, the chirp of
the emitted high harmonic light and other quantities of interest in
atomic systems\cite{lewenstein,kazamias04}.  However, as was shown in
\cite{walters08}, the 
plane wave approximation is not adequate to describe the returning
electron because of the tremendous distortion caused by the
electron's interaction with the ionic potential and by exchange
effects with the other electrons in the molecule.\cite{Le08,Smirnova08}  For the
time-reversed problem of photoionization, it is known that the plane
wave approximation is prone to error for photoelectron energies
smaller than the deepest K-shell binding energy. Energies attained in
high harmonic generation experiments usually fall below this range.


This section gives a semiclassical model of the free electron
propagation which improves upon the Corkum/Lewenstein model by
connecting with calculated short-range wavefunctions.  
Any method could be used to calculate
these short-range wavefunctions.  In this treatment, the tunneling
wavefunction is modeled semiclassically using ideas based on the
initial value representation \cite{miller2001,nakamura2005}.  The
continuum wavepacket of the recolliding electron is described in terms of
electron-molecule scattering states as used in \cite{walters08}, 
calculated in the absence of an
external electric field.

In this way, the molecular potential nontrivially affects the electron wavefunction
at the two times when the electron is near the molecular ion.  When the
electron is far from the molecule, the comparatively simple evolution
of its wavefunction is described using the shortest-time, dominant
contribution to the Gutzwiller propagator \cite{gutzwiller}.  Finally,
stationary phase arguments serve to identify the isolated trajectories
that encapsulate the effect of the electron propagation in the field
on high harmonic generation, greatly reducing the computational burden
of propagating the continuum electron wavefunction.

\subsection{Tunneling Ionization}
\begin{figure}
\begin{center}
\includegraphics[width=3.375in]{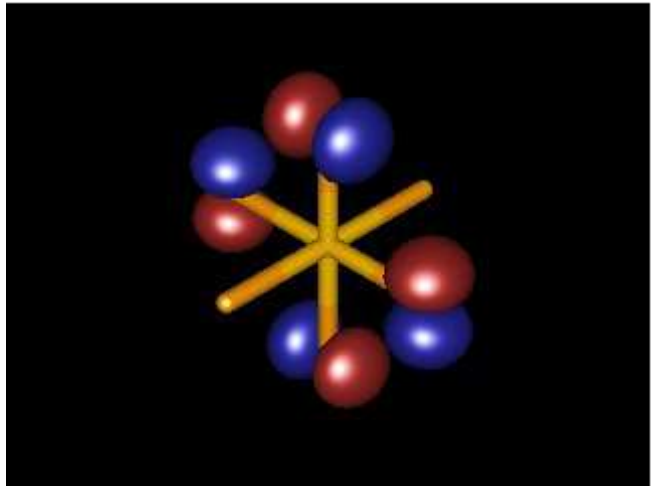}
\end{center}
\caption{One of three degenerate orbitals of
  SF$_{6}$. Red denotes positive lobes; blue denotes negative lobes.}
\label{fig:sf6_homo}
\end{figure}

During the ionization step, the tunneling electron wavefunction is
described in a simple 1-D WKB tunneling picture, in which electrons
are allowed to tunnel only in directions parallel to the laser's
applied electric field.  This approach is motivated by the
semiclassical ``initial value representation''
\cite{miller2001,nakamura2005}, where a source 
wavefunction acquires an imaginary phase (and hence an exponential growth
or decay) along a trajectory that passes through a classically forbidden
region.  The unperturbed highest occupied molecular orbital (HOMO)
here serves as the source 
wavefunction, so that the tunneling wavefunction is approximated by
the unperturbed HOMO in the classically-allowed region near the
molecule, connecting to a WKB exponential which is set equal to the HOMO at the
inner turning point and decays exponentially until it reaches
the outer turning point.  SF$_{6}$ has three degenerate HOMOs: one of
these is illustrated in Figure \ref{fig:sf6_homo}.

\begin{figure}
\begin{center}
\includegraphics[width=3.375in]{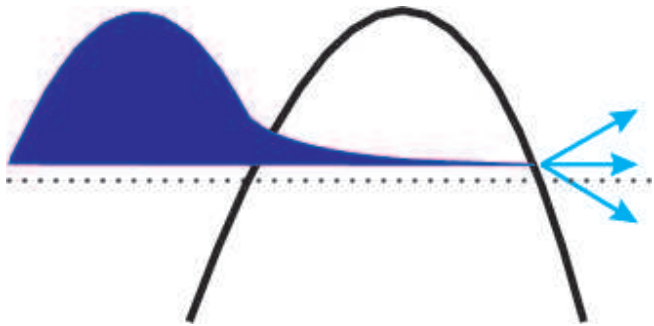}
\end{center}
\caption{The tunneling wavefunction is approximated as
  an unperturbed molecular HOMO inside the classically allowed region,
  connecting to a decreasing WKB exponential in the classically
  forbidden region.  Stationary phase trajectories leave from the
  outer turning point, beginning with zero velocity at time of ionization.}
\label{fig:tunnelinggraphic}
\end{figure}

The classically forbidden region, illustrated in Figure
\ref{fig:tunnelinggraphic}, is defined by 
two turning points, $a$ and $b$, between which $V(x)-E>0$.  The slope of
$V(x)$ is $C_{1}$ at inner turning point $a$ and $C_{2}$ at outer
turning point $b$.  The tunneling wavefunction $\psi_{t}(\vec{r},t)$
is now found by applying WKB connection formulas.  The appendix
derives the ratio of the tunneling
wavefunction at the outer turning point to the tunneling wavefunction
at the inner turning point 
\begin{equation}
\frac{\psi(x=b)}{\psi(x=a)}=\frac{1}{2}e^{-\Gamma}|\frac{C_{1}}{C_{2}}|^{1/6}
\frac{\text{Bi}(0)}{\text{Ai}(0)}
\end{equation}

In this approximation, the tunneling wavefunction behaves like an Airy
$Bi$ function near the outer turning point, i.e. it has no linear
complex phase term.  This property will be revisited in the section
dealing with stationary phase analysis.

\subsection{Semiclassical Propagation}
When the active electron has tunneled free from the molecule, the
evolution of its wavefunction is controlled by the oscillating electric
field of the laser, plus a residual Coulomb attraction to the
molecular ion.  This relatively simple evolution continues until the
electron returns to the molecule, when the complicated molecular
potential again becomes significant.  During this excursion, until it
re-enters the non-Coulomb part of the potential, the continuum
wavefunction can be approximated using Gutzwiller's semiclassical
propagator\cite{gutzwiller}:
\begin{equation}
\begin{split}
K(\vec{r},t;\vec{r}_{0},t_{0})=\sum_{\text{cl. traj.}} 
(2\pi i)^{-3/2}\sqrt{C(\vec{r},t;\vec{r}_{0},t_{0})} \times \\
\exp(iS(\vec{r},t;\vec{r}_{0},t_{0})-i\phi)
\label{eq:gutzwiller}
\end{split}
\end{equation}

Here $S(\vec{r},t;\vec{r}_{0},t_{0})$ is the action integral 
$S=\int L(q,\dot{q},t)dt$ calculated for a classical trajectory
starting at $(\vec{r}_{0},t_{0})$ and ending at $(\vec{r},t)$, while 
$C(\vec{r},t;\vec{r}_{0},t_{0})=|\frac{-\partial^{2}S}{\partial
  r_{0,A}} \partial r_{B}|$, where $r_{A}$ is the A-component of the
vector $\vec{r}$.  $\phi$ is a phase factor equal to $\frac{\pi}{2}$
times the number of conjugate points crossed by the
trajectory\cite{gutzwiller}. 

This propagator acts on the tunneling wavepacket $\psi_{t}$ to give a
semiclassical continuum wavepacket 
\begin{equation}
\psi_{c}(\vec{r},t)=\int d^{3}\vec{r}_{0} \int dt_{0}
K(\vec{r},t;\vec{r}_{0},t_{0})\psi_{t}(\vec{r}_{0},t_{0}).
\label{eq:psic}
\end{equation}
until the molecular potential asserts itself during the rescattering.

During the terminal portion of the scattering process, when the
scattering wavefunction acquires its maximum
dipole matrix element with the molecular
HOMO into which it recombines,  the electronic wavefunction is
expanded into a truncated (but in principle complete) basis of
field-free electron-molecule scattering orbitals, calculated using
techniques described in references 
\cite{tonzani2005,tonzani2006a,tonzani07}.  Beyond the range of the
molecular potential, i.e. for $r>r_{0}$, the $(l,m)$-th independent
scattering state is expressed as a partial wave expansion in terms of
incoming and outgoing Coulomb radial functions $f^{\pm}_{El}(r)$ and
the scattering S-matrix as
\begin{equation}
\begin{split}
\psi_{E,lm}(\vec{r})=\frac{1}{i
  \sqrt{2}}f^{-}_{El}(r)Y_{lm}(\theta,\phi)- \\
\frac{1}{i \sqrt{2}}\sum_{l^{\prime}m^{\prime}}
  f^{+}_{El^{\prime}}(r)Y_{l^{\prime}m^{\prime}}(\theta,\phi)
  S_{l^{\prime}m^{\prime};lm}(E), r \geq r_{0}.
\end{split}
\end{equation}
The laser electric field is typically far smaller when the electron
returns to the ion than it was when it departed.  Even if this were
not the case, this external force is less than that due to the
electron-ion interaction force when the electron is in the molecular
field.  Neglecting the effect of the external field on the electron
during its brief recollision with the ion, the time-dependent
wavefunction becomes 
\begin{equation}
\psi_{\text{scat}}(\vec{r},t)=\int dE \sum_{lm} A_{lm}(E)
\psi_{E,lm}(\vec{r})e^{-i E t} 
\end{equation}
and the expansion coefficients $A_{l,m}(E)$ are given by
\begin{equation}
A_{lm}(E)=e^{iEt}\int
d^{3}\vec{r}\psi^{*}_{E,lm}(\vec{r})\psi_{c}(\vec{r},t)
\label{eq:alm}
\end{equation}
for some chosen time $t$ when $\psi_{c}$ is projected onto the
scattering states.  This projection time is chosen so that the bulk of
the returning wavepacket is approaching close to the ion but has not
yet recollided.

The dipole recombination amplitudes
\begin{equation}
d_{l,m}(E)=\bra{\psi_{E,lm}}\hat{\epsilon} \cdot \vec{r}
  \ket{\psi_{\text{HOMO}}}
\end{equation}
are calculated between the distorted scattering states and the
molecular HOMO into which the electron recombines.

Equations [\ref{eq:psic}] and [\ref{eq:alm}] now define the expansion
coefficients $A_{lm}(E)$ for some chosen projection time $t$, in terms
of two three-dimensional integrals over initial and final positions,
one integral over the time of ionization, and a summation over all
possible classical trajectories.  Stationary phase arguments
dramatically simplify the calculation of these expansion coefficients.

\subsection{Stationary phase calculation of scattering coefficients}

A difficulty which must be resolved when using the Gutzwiller
propagator to describe the evolution of the continuum wavefunction far
from the molecule and scattering/tunneling wavefunctions to describe
its evolution near the molecule is the very different physical
pictures employed in the two treatments.  The Gutzwiller propagator
$K(\vec{r},t;\vec{r}_{0},t_{0})$ defined in 
equation [\ref{eq:gutzwiller}] involves a summation
over the classical paths having $(\vec{r}_{0},t_{0})$ and
$(\vec{r},t)$ as their endpoints, whereas short-range treatments of
scattering and tunneling use more familiar wavefunction descriptions.
Moreover, previous semiclassical treatments such as
\cite{corkum93,lewenstein94} have found only a few such
classical paths to be significant in describing the high harmonic process.

Both of these difficulties may be resolved by using stationary phase
techniques to search for families of classical trajectories which add
together to give nonvanishing contributions to the expansion
coefficients $A_{l,m}(E)$ defined in equation \ref{eq:alm}.

Consider the contribution to $A_{l,m}(E)$ made by all classical paths
originating at a given point $(\vec{r}_{0},t_{0})$.  Making the usual
assumption of isolated trajectories, there will be one classical path
connecting starting point $(\vec{r}_{0},t_{0})$ to any given ending point
$(\vec{r},t)$, which will have the property of extremizing the action
integral $S(\vec{r},t;\vec{r}_{0},t_{0})=\int L(q,\dot{q},t) dt$ with
respect to any perturbation which does not alter the starting or
finishing points.  The connection to the wavefunction treatment of the
scattering and tunneling wavefunctions is made by noting that
$S(\vec{r},t;\vec{r}_{0},t_{0})$ appears as a phase term in the
Gutzwiller propagator.  Thus, the change in accumulated action which
is caused by changing the endpoints of a classical path corresponds to
a change in the phase of the propagated wavefunction.

This variation of the action integral with respect to the starting and
ending points of a trajectory, is known (see Goldstein et al
\cite{goldstein} section 8.6) as the $\Delta$ variation of the
action, given by
\begin{equation}
\Delta S(\vec{r},t;\vec{r}_{0},t_{0})=(p_{A}\delta q_{A}-\hat{H}(t)\delta t)|
_{\text{initial}}^{\text{final}}.
\label{eq:deltavariation}
\end{equation}

The condition for a nonoscillating integrand in Eq. \ref{eq:alm} is
now that the variation of the phase of the Gutzwiller propagator
resulting from Eq. \ref{eq:deltavariation} must be offset by the
variation of phase of the tunneling or scattering wavefunction.
In this paper, a trajectory
where the $\Delta$ variation of the action is counterbalanced by the
phase of the tunneling wavefunction at $(\vec{r}_{0},t_{0})$ and by
the phase of the scattering wavefunction at $(\vec{r},t)$ will be
known as a
``stationary phase trajectory.''  The initial and final points
$(\vec{r}_{0},t_{0})$ and $(\vec{r},t)$
identify the points in the 7-dimensional integral of Eq. \ref{eq:alm} where the
integrand oscillates slowly, giving a non-canceling contribution to
the expansion coefficients $A_{lm}(E)$.

Stationary phase trajectories will be found to correspond with the
trajectories used in prior semiclassical theories of high harmonic
generation.  However, the current approach allows for more detailed
treatments of the short-range tunneling and scattering wavefunctions,
where semiclassical methods may give unsatisfactory descriptions of
the physics.

Stationary phase trajectories may be found by expanding the
phase-oscillating parts of the 7D integral from Eq. \ref{eq:alm}
\begin{equation}
\begin{split}
A_{lm}(E)=\int d^{3}\vec{r} \int d^{3}\vec{r}_{0} \int dt_{0} e^{iEt}
\sqrt{C(\vec{r},t;\vec{r}_{0},t_{0})} \\ 
\exp(i S(\vec{r},t;\vec{r}_{0},t_{0})-i\phi)
\psi^{*}_{E,lm}(\vec{r},t)\psi_{t}(\vec{r}_{0},t_{0}).
\end{split}
\end{equation}
about the starting point $(\vec{r}_{0c},t_{0c})$ and about the ending point
$(\vec{r}_{c},t)$

The $\Delta$ variation gives the expansion of the action integral
\begin{equation}
\begin{split}
S(\vec{r}_{c}+\delta \vec{r},t;\vec{r}_{0c}+\delta
\vec{r}_{0},t_{0c}+\delta t_{0})=
S(\vec{r}_{c},t;\vec{r}_{0c},t_{0c})+ \\
p_{A}\delta r_{A} 
+\frac{1}{2} \frac{\partial^{2} S}{\partial r_{A}\partial r_{B}}\delta
r_{A} \delta r_{B}+
p_{0A} \delta r_{0A}+ \\
\frac{1}{2} \frac{\partial^{2} S}{\partial r_{0A} \partial r_{0B}}
\delta r_{0A} \delta r_{0B} 
+\hat{H} \delta t_{0}
+\frac{1}{2}\frac{\partial^2 R}{\partial t_{0}^2} (\delta t_{0})^2.
\end{split}
\end{equation}


The condition for a nonoscillatory integrand is now that the
first-order terms in $\delta \vec{r}_{0}$,$\delta \vec{r}$ and $\delta
t_{0}$ must disappear.

For the tunneling wavefunction, which resembles a declining WKB
exponential in the forbidden region and has no oscillatory component,
this corresponds to a trajectory which leaves the molecule with zero
initial momentum.

For the scattering wavefunction, ignoring the angular derivative of
the spherical harmonics, the phase evolution of the scattering states
is given by the asymptotic form of the Coulomb wave functions
$f^{\pm}_{El}(r)$ 
\begin{equation}
\begin{split}
\psi^{*}_{E,lm}(\vec{r}_{c}+\delta \vec{r})=
\frac{(2 \pi i)^{-3/2}}{-i \sqrt{2}} f^{-*}_{El}(r_{c})
Y^{*}_{lm}(\theta_{c},\phi_{c}) \exp(i k(r_{c})\delta r) - \\
\sum_{l^{\prime},m^{\prime}}\frac{1}{-i \sqrt{2}}
f^{+}_{El^{\prime}} Y^{*}_{l^{\prime}m^{\prime}}(\theta_{c},\phi_{c}) 
S_{l^{\prime}m^{\prime};lm}(E) \exp(-i k(r_{c}) \delta r))
\end{split}
\end{equation}
 where $k_{l}(r)=\sqrt(2(E-V_{l}(r)))$.  The condition for a
 nonoscillatory phase integrand is that
 $p_{r}=-k_{l}(r)=\sqrt(2(E-V_{l}(r)))$.

Neglecting the angular
derivatives of the spherical harmonics is justified because the action
along the recollision trajectory is much greater than that in a
low-order angular solution. It translates into a stationary phase
condition that the trajectory must
return with zero angular momentum.

Finally, setting the coefficients of $\delta t_{0}$ to zero yields the
condition that 
\begin{equation}
\frac{\vec{p}_{0}^{2}}{2m}+V(\vec{r}_{0},t_{0})=E_{\text{HOMO}}.
\end{equation}


Thus, a stationary phase trajectory is launched with zero momentum
from the classical turning point and returns to the molecule with zero
angular momentum, and with kinetic energy equal to the energy of the
scattering state.  This is a familiar result from, e.g., \cite{corkum93,
lewenstein94}, with the distinction that the present work considers the
effect of the electron-ion Coulomb interaction during the continuum
propagation of the 
electron.  Also, the wavefunction is projected onto scattering states shortly
before recollision, rather than treating the electron in a Volkov
approximation throughout the recollision with the molecular ion.  Because the
electron-molecule scattering states are calculated with no external
electric field present, there is a slight dependence on the time at which the
wavefunction is projected onto scattering states; here the projection
is made when $\omega t$ for the laser cycle is equal to 3.9, i.e. when
the short trajectories with energies equal to the energy of the 39th harmonic
have nearly returned to the molecule.

Because the linear phase variation of the integrand vanishes in the
vicinity of a stationary phase trajectory, the expansion coefficients
$A_{lm}(E)$ are now found via Gaussian integrals:
\begin{equation}
\begin{split}
A_{lm}(E)=(2 \pi i)^{-3/2}\frac{i}{\sqrt{2}} f^{-*}_{El}(r_{c}) 
Y^{*}_{lm}(\theta_{c},\phi_{c}) \times\\ 
\exp(iS(\vec{r}_c,t;\vec{r}_{0c},t_{0c})-i\phi) \psi_{t}(\vec{r}_{0c})e^{-i
  E_{\text{HOMO}} t_{0c}}  \times\\
 ((I)=\int d(\delta t_{0}) \exp(\frac{i}{2}\frac{\partial^{2} S}{\partial
  t_{0}^{2}} (\delta t_{0})^{2})) \times \\
((II)=\int d^{3} (\delta \vec{r}_{0}) \exp(\frac{i}{2} \frac{\partial^2
  S}{\partial r_{0A} \partial r_{0B}} \delta r_{0a} \delta r_{0B})) \times \\
((III)=\int d^{3} (\delta \vec{r}) \exp(\frac{i}{2} \frac{\partial^2
  S}{\partial r_{A} \partial r_{B}} \delta r_{A} \delta r_{B}))
\end{split}
\end{equation}

where the integrals labeled $(I)$,$(II)$ and $(III)$ are evaluated as
\begin{eqnarray}
(I) = \sqrt{2 \pi i}|\frac{\partial^{2} S}{\partial t_{0}^{2}}|^{-1/2}
  \\
(II) = (2 \pi i)^{3/2}|\frac{\partial^{2} S}{\partial r_{0A} 
\partial r_{0B}}|^{-1/2} \\
(III) = (2 \pi i)^{3/2}|\frac{\partial^{2} S}{\partial r_{A} 
\partial r_{B}}|^{-1/2}
\end{eqnarray}
yielding expansion coefficients
\begin{equation}
\begin{split}
A_{lm}(E)=(2 \pi i)^{2} \sqrt{|\frac{\partial r_{A}}{\partial
    p_{0B}}|} \times\\
(\frac{\partial^2 S}{\partial t_{0}^{2}})^{-1/2}
    \frac{i}{\sqrt{2}}  \times\\
f^{-*}_{El}(r_{c}) 
Y^{*}_{lm}(\theta_{c},\phi_{c}) \exp(i
S(\vec{r}_c,t;\vec{r}_{0c},t_{0c})-i\phi) \times\\
\psi_{t}(\vec{r}_{0c})e^{-i
  E_{\text{HOMO}} t_{0c}}
\end{split}
\end{equation}
Once these expansion coefficients have been calculated, the dipole
matrix element between the distorted scattering wave and the molecular
HOMO is simply
\begin{equation}
\vec{D}(E)=\sum_{lm} A_{lm}(E) \vec{d}_{lm}(E).
\end{equation}

\section{Comparison with Experiment}

The model of vibrational interference connects with this treatment of
the high harmonic process when $\hat{I}\hat{\vec{R}}$ is set to
$\vec{D}(E)$.  This is broken up by setting
$\hat{I}=\psi_{t}(\vec{r}_{0},t_{0})$ and $\hat{\vec{R}}=\vec{D}(E)/\hat{I}$.
Both of these
quantities are calculated for a molecule at the equilibrium geometry,
and for a molecule displaced by 0.1 bohr in the normal mode
coordinate.  This involves recalculating the scattering states and
recombination dipoles for each distorted molecular geometry.

The modulation of the 39th harmonic was chosen for purposes of
comparison with experiment, since this harmonic was considered in detail in
\cite{wagner2006}.  The 39th harmonic falls close to the measured
cutoff, and can only be produced by a half-cycle coming close to the
maximum of the gaussian envelope of the laser pulse.  

The JILA experiment used a gas jet as a source of SF$_{6}$, giving no
preferred molecular orientation.  
However, both ionization and recombination amplitudes are highly
dependent on orientation.  Therefore, a rotational average was calculated for
both the static and oscillatory parts of the harmonic intensity.  Only
those polarizations perpendicular to the propagating laser beam for a
given molecular orientation were included in these averages.

\begin{figure}
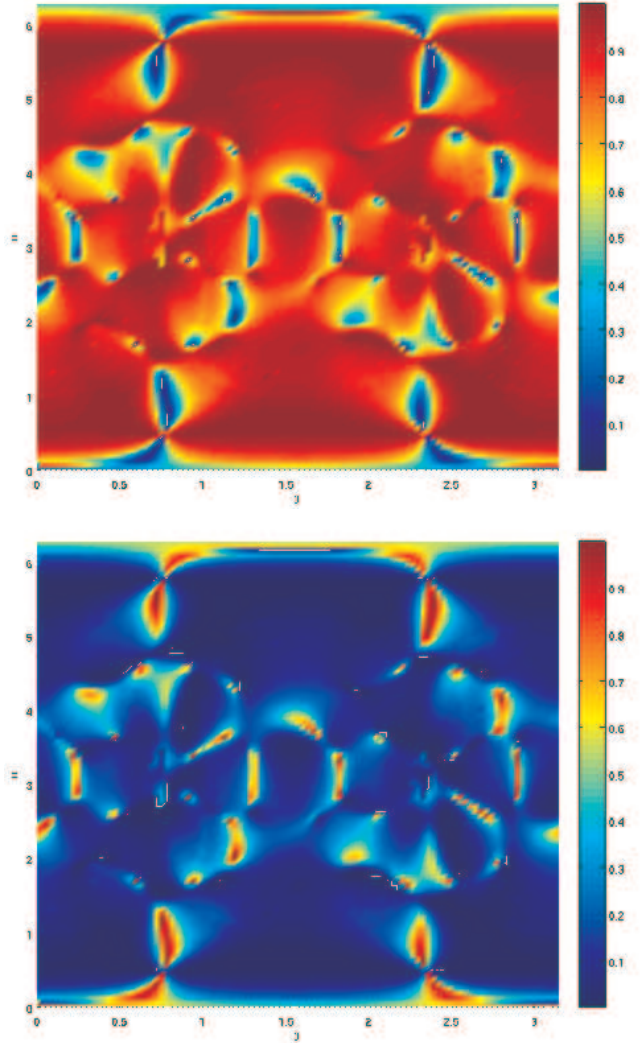

\begin{center}
\includegraphics[width=3.375in]{t2g1.popfrac.0.eps2}
\includegraphics[width=3.375in]{t2g1.popfrac.1.eps2}
\end{center}
\caption{Population of $\ket{0}$ and $\ket{1}$ vibrational states after
  the high harmonic process as a
  function of angle for the $T_{2g}$ normal mode transforming like $xy$.}
\label{fig:t2gpopfrac}
\end{figure}

\begin{figure}
\begin{center}
\includegraphics[width=3.375in]{t2g1.popmod.0.real.eps2}
\includegraphics[width=3.375in]{t2g1.popmod.0.imag.eps2}
\end{center}
\caption{Modulation of HHG signal resulting in final vibrational state
  $\ket{0}$ as a
  function of angle for the $T_{2g}$ normal mode transforming like
  $xy$. a) real component of modulation b) imaginary component of modulation}
\label{fig:t2g0popmod}
\end{figure}

\begin{figure}
\begin{center}
\includegraphics[width=3.375in]{t2g1.popmod.1.real.eps2}
\includegraphics[width=3.375in]{t2g1.popmod.1.imag.eps2}
\end{center}
\caption{Modulation of HHG signal resulting in final vibrational state
  $\ket{1}$ as a
  function of angle for the $T_{2g}$ normal mode transforming like
  $xy$. a) real component of modulation b) imaginary component of modulation}
\label{fig:t2g1popmod}
\end{figure}

\begin{figure}
\begin{center}
\includegraphics[width=3.375in]{eg1.popfrac.0.eps2}
\includegraphics[width=3.375in]{eg1.popfrac.1.eps2}
\end{center}
\caption{Population of $\ket{0}$ and $\ket{1}$ vibrational states after
  the high harmonic process as a
  function of angle for the $E_{g}$ normal mode transforming like 
  $2z^{2}-x^{2}-y^{2}$. }
\label{fig:egpopfrac}
\end{figure}

\begin{figure}
\begin{center}
\includegraphics[width=3.375in]{eg1.popmod.0.real.eps2}
\includegraphics[width=3.375in]{eg1.popmod.0.imag.eps2}
\end{center}
\caption{Modulation of HHG signal resulting in final vibrational state
  $\ket{0}$ as a
  function of angle for the $E_{g}$ normal mode transforming like
  $2z^{2}-x^{2}-y^{2}$. a) real component of modulation b) imaginary component of modulation}
\label{fig:eg0popmod}
\end{figure}

\begin{figure}
\begin{center}
\includegraphics[width=3.375in]{eg1.popmod.1.real.eps2}
\includegraphics[width=3.375in]{eg1.popmod.1.imag.eps2}
\end{center}
\caption{Modulation of HHG signal resulting in final vibrational state
  $\ket{1}$ as a
  function of angle for the $E_{g}$ normal mode transforming like
  $2z^{2}-x^{2}-y^{2}$. a) real component of modulation b) imaginary
  component of modulation.}
\label{fig:eg1popmod}
\end{figure}

\begin{figure}
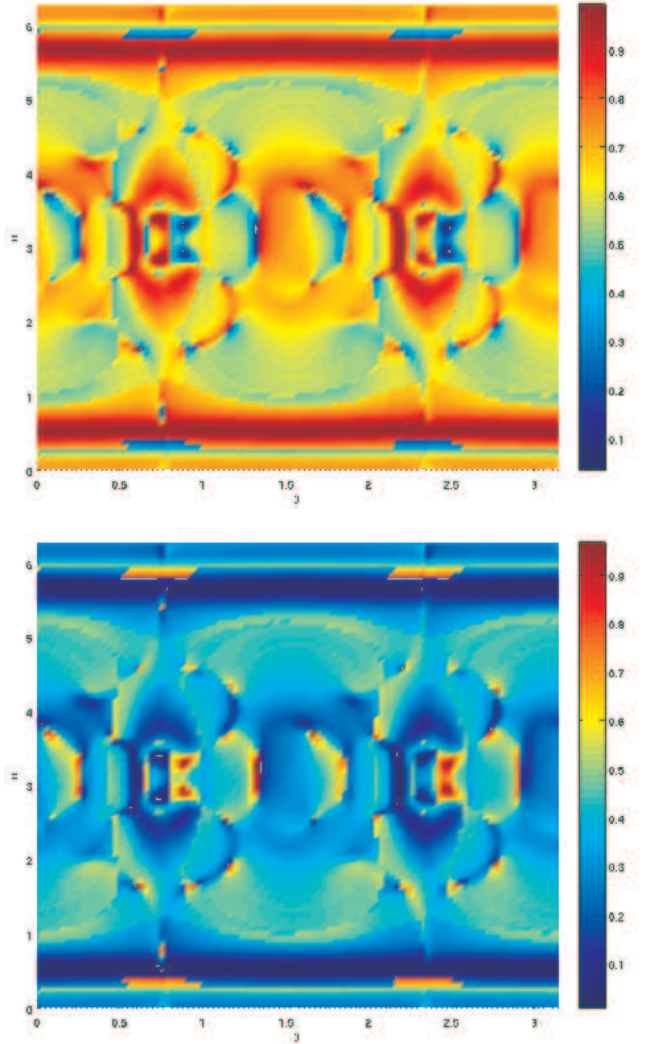

\begin{center}
\includegraphics[width=3.375in]{a1g.popfrac.0.eps2}
\includegraphics[width=3.375in]{a1g.popfrac.1.eps2}
\end{center}
\caption{Population of $\ket{0}$ and $\ket{1}$ vibrational states after
  the high harmonic process as a
  function of angle for the totally symmetric $A_{1g}$ mode.}
\label{fig:a1gpopfrac}
\end{figure}

\begin{figure}
\begin{center}
\includegraphics[width=3.375in]{a1g.popmod.0.real.eps2}
\includegraphics[width=3.375in]{a1g.popmod.0.imag.eps2}
\end{center}
\caption{Modulation of HHG signal resulting in final vibrational state
  $\ket{0}$ as a
  function of angle for the totally symmetric $A_{1g}$ mode. a) real component of modulation b) imaginary component of modulation}
\label{fig:a1g0popmod}
\end{figure}

\begin{figure}
\begin{center}
\includegraphics[width=3.375in]{a1g.popmod.1.real.eps2}
\includegraphics[width=3.375in]{a1g.popmod.1.imag.eps2}
\end{center}
\caption{Modulation of HHG signal resulting in final vibrational state
  $\ket{1}$ as a
  function of angle for the totally symmetric $A_{1g}$ mode. a) real component of modulation b) imaginary
  component of modulation.}
\label{fig:a1g1popmod}
\end{figure}


A previous work\cite{walters2007a} compared only the angular averaged
modulation of the entire signal to experiment, by calculating the
modulation to first order in raising and lowering operators.  For the
present work, the modulation at a particular molecular orientation
was calculated using the full transfer matrices $\underline{I}$,
$\underline{R}$ and $\underline{M}$.  The resulting vibrational state
populations and modulations of the different components of the HHG
signal show a rich angular structure which is lost upon angular
averaging. 

Note that all calculations presented here have used the separable
approximation of Eq.\ref{eq:separability} for the amplitudes of the different
vibrational modes.  This is expected to be accurate to the extent that
the ground vibrational state dominates, meaning that for the
amplitudes that we have calculated here, the higher-order nonseparable
pathways are at least beginning to become important, which diminishes
the validity of this approximation.  Nevertheless, it is still
expected to have at least qualitative and perhaps even
semi-quantitative validity for the range of parameters studied here.   

Using the separable approximation, the final populations of the
$\ket{0}$ and $\ket{1}$  for each normal mode and their 
modulation as a function of time were calculated
as a function of angle using equation \ref{eq:intensity}.  Figures
\ref{fig:t2gpopfrac},\ref{fig:t2g0popmod},\ref{fig:t2g1popmod} show
final state populations and the real and imaginary components of the
modulation as a function of angle for the $T_{2g}$ mode transforming
like $xy$.  Figures
\ref{fig:egpopfrac},\ref{fig:eg0popmod},\ref{fig:eg1popmod} show final
state populations and modulations for the $E_{g}$ mode transforming
like $2z^{2}-x^{2}-y^{2}$, while Figures
\ref{fig:a1gpopfrac},\ref{fig:a1g0popmod},\ref{fig:a1g1popmod} show
final populations and modulation fractions for the totally symmetric
$A_{1g}$ mode.  In all three modes, the population of the $\ket{1}$
vibrational state is modulated much more heavily than the population
of the $\ket{0}$ state.

A noteworthy feature of these figures is that in regions where Raman
excitation during the HHG pulse is weak, the modulation of the HHG
signal due to $d_{0}^{*} d_{0}$ (ie, HHG processes ending with the
molecule in the ground vibrational state) tends to cancel modulation
due to $d_{1}^{*} d_{1}$ (leaving the molecule in the first
vibrational state).  For this reason, angle-averaged modulations of
the overall HHG signal due to the T$_{2g}$ and E$_{g}$ normal modes
give almost zero overall modulation, while the modulations of the
$d_{0}^{*} d_{0}$ or $d_{1}^{*} d_{1}$ give a large fractional
modulation. (The A$_{1g}$ mode experiences strong Raman excitation at
all molecular orientations, and experiences less cancellation as a
result.)

It is not clear why this cancellation is not observed in the
experiment, where the T$_{2g}$ mode is typically the most visible
\cite{wagnerpersonal}.  This could arise due to a variety of causes,
such as preferential detection of the  $d_{1}^{*} d_{1}$ component of
the signal relative to the $d_{0}^{*} d_{0}$ component, or some
mechanism changing the phase between the modulation of the two
components and thereby eliminate the cancellation. Table
\ref{table:01popfrac} shows the fraction of molecules finishing the
HHG process in the $\ket{0}$ and $\ket{1}$ states, while table
\ref{table:01popmod} compares the modulation of the $d_{0}^{*} d_{0}$
and $d_{1}^{*} d_{1}$ components of the signal, the modulation of the
total signal, and the two experimental runs which were able to detect
modulations at all three vibrational frequencies.  Tables
\ref{table:manystatepopfrac} and \ref{table:manystatepopmod}  shows the same information when the HHG
process is allowed to populate vibrational states up to $\ket{4}.$
It is apparent from Table \ref{table:manystatepopfrac} that including higher
vibrational states of the T$_{2g}$ and E$_{g}$ modes do not greatly
affect the calculated modulations.  For the $A_{1g}$ mode, which
undergoes stronger Raman excitation, the presence of $\ket{2}$ and
higher vibrational states does become important.

\begin{table}[ht]
\caption{Peak-to-peak Modulation, Theory vs. Experiment (2 state
  model)}
\centering
\begin{tabular}{cccccc}
\hline\hline
Mode & Experiment 1 & Experiment 2 & $\ket{0}$ & $\ket{1}$ & Total Signal \\
A$_{1g}$ & .06 & .105 & .0439 & .187 & .0246 \\
T$_{2g}$ & .105 & .122 & .0010 & .236 & .0016 \\
E$_{g}$ & .025 & .029 & .0255 & .121 & .0020 \\[1ex]
\hline
\end{tabular}
\label{table:01popmod}
\end{table}

\begin{table}[ht]
\caption{Peak-to-peak Modulation, Theory vs. Experiment (5 state
  model)}
\centering
\begin{tabular}{ccccccc}
\hline\hline
Mode & $\ket{0}$ & $\ket{1}$ & $\ket{2}$ & $\ket{3}$ & $\ket{4}$ & Total
modulation \\
A$_{1g}$ & .0495 & .249 & .508 & 1.01 & 1.44 & .0901 \\
T$_{2g}$ & .0099 & .232 & .708 & 1.45 & 1.84 & .0011 \\
E$_{g}$ & .0264 & .142 & .907 & 1.55 & 1.88 & .0071 \\[1ex]
\hline
\end{tabular}
\label{table:manystatepopmod}
\end{table}

\begin{table}[ht]
\caption{Vibrational State Population After HHG process (2 state model)}
\centering
\begin{tabular}{ccc}
\hline\hline
Mode &$\ket{0}$ pop. & $\ket{1}$ pop.\\
A$_{1g}$ & 0.71 & 0.29 \\
T$_{2g}$ & 0.87 & 0.13 \\
E$_{g}$ & 0.69 & 0.33 \\[1ex]
\hline
\end{tabular}
\label{table:01popfrac}
\end{table}

\begin{table}[ht]
\caption{Vibrational State Population After HHG process (5 state model)}
\centering
\begin{tabular}{cccccc}
\hline\hline
Mode &$\ket{0}$ pop. & $\ket{1}$ pop. & $\ket{2}$ pop. & $\ket{3}$
pop. & $\ket{4}$ pop. \\
A$_{1g}$ & .611 & .310 & .0719 & 7.03$\times10^{-3}$ & 4.37$\times10^{-4}$\\
T$_{2g}$ & .955 & .0438 & 6.92$\times10^{-4}$ & 7.90$\times10^{-6}$ & 8.73$\times10^{-8}$ \\
E$_{g}$ & .826 & .168 & 5.91$\times10^{-3}$ & 1.63$\times10^{-4}$ & 3.59$\times10^{-6}$ \\[1ex]
\hline
\end{tabular}
\label{table:manystatepopfrac}
\end{table}

Although the agreement with experiment is not perfect, it is
nevertheless significant that the simple model of vibrational
interference presented here agrees with experiment to the correct
order of magnitude.  This is particularly notable in light of
conventional Raman spectroscopy, in which the A$_{1g}$ peak is 20
times more prominent than the others.
It is difficult to precisely gauge the agreement
of theory and experiment, due to the paucity of experimental data.
Peak-to-peak modulations vary extensively from one experimental run to
another \cite{wagnerpersonal} 
with this SF$_{6}$ experiment.  The modulation at 525 $cm^{-1}$,
corresponding to the T$_{2g}$ 
mode, appears most prominently in the experimental data, yet it gives the
smallest modulation in this treatment.  The prominence of the T$_{2g}$
mode modulation may suggest that the off-diagonal Jahn-Teller coupling
$V_{T_{2g}}$ is larger than obtained in the present calculations.
Alternatively, it may be necessary to model the
experiment in more detail -- i.e. to incorporate the spatially-varying
laser intensity, the
uncontrolled carrier envelope phase, the combination of multiple laser
half cycles, etc 
-- beyond that which has been included in the present theoretical description.

\section{Conclusions}
The problem of high harmonic generation in molecules can be
conceptually separated into two parts: the evolution of the continuum
electron, and the evolution of the internal (vibrational) wavefunction
of the parent ion.  This paper describes the evolution of the
continuum electron in a model which combines a semiclassical treatment
of the propagation with a fully quantum mechanical description of the
electron-molecule scattering.  This flexible and robust model has a
simple conceptual link to existing semiclassical models, yet it allows
for a sophisticated treatment of the complicated electron-molecule
scattering.  The internal dynamics of the parent ion are tracked
throughout the high harmonic process.  Together, these two innovations
serve to give an unprecedented view of high harmonic generation in a
comparatively large, complicated molecule with many internal degrees
of freedom, 
giving results which agree with experiment to within an order of magnitude.

The possibility that high harmonic generation may serve as an ultrafast
interferometric probe of a molecular vibrational wavefunction is extremely
promising.  Such a wavefunction need not be prepared by an
initial Raman pulse, as was the case for the JILA experiment.  Instead, a
preparatory pulse could photoionize a molecule, excite it to a
higher electronic state, or trigger the beginning of some other
chemical process.  In this way, vibrational wavepacket evolution
during chemical processes could be observed as it happens.

\section{Acknowledgments}
We thank the group of H. Kapteyn and M. Murnane for helpful
discussions. This work was supported in part by the Office of Science,
Department of Energy, and in part by the NSF EUV Engineering Research
Center. 

\section{Appendix: Tunneling Ionization}
In this approximate treatment, the wavefunction in the forbidden region is
found using the WKB connection formulas.  As is illustrated in Figure
\ref{fig:tunnelinggraphic}, the classically forbidden region is defined by
two turning points, $a$ and $b$, aligned in the downfield direction,
between which $V(x)-E>0$.  The slope of 
$V(x)$ is $C_{1}>0$ at the inner turning point $a$ and $C_{2}<0$ at the outer
turning point $b$.

Near turning point $a$, $k^{2}(x)\equiv2m(E-V(x))\approx C_{1}(a-x)$ and
the time independent Schr\"odinger equation is 
\begin{equation}
\psi^{\prime \prime}+k^{2}(x)\psi=0,
\end{equation}
which has solutions near $x=a$ of
\begin{equation}
\psi(x)=\text{Ai}(\frac{C_{1}(x-a)}{C_{1}^{2/3}}) b_{1} +
\text{Bi}(\frac{C_{1}(x-a)}{C_{1}^{2/3}}) b_{2}
\end{equation}
where $\text{Ai}$ and $\text{Bi}$ are Airy functions, asymptotically behaving like
\begin{eqnarray}
\text{Ai}(z)  \Rightarrow_{z \rightarrow \infty} & (2 \pi)^{-1/2} z^{-1/4}
\exp[\frac{-2}{3} z^{3/2}] \\
\text{Ai}(z)  \Rightarrow_{z \rightarrow -\infty} & \pi^{-1/2} (-z)^{-1/4}
\sin(\frac{2}{3}(-z)^{3/2}+\pi/4) \\
\text{Bi}(z)  \Rightarrow_{z \rightarrow \infty} & \pi^{-1/2} z^{-1/4}
\exp(\frac{2}{3} z^{3/2}) \\
\text{Bi}(z)  \Rightarrow_{z \rightarrow -\infty} & \pi^{-1/2} (-z)^{-1/4}
\cos(\frac{2}{3} (-z)^{3/2}+\pi/4)
\end{eqnarray}
where $z=C_{1}^{1/3}(x-a)$

Similarly, near $x=b$
\begin{equation}
\psi(x)=\text{Ai}(C_{2}^{1/3}(b-x)) d_{1}+\text{Bi}(C_{2}^{1/3}(b-x)) d_{2}
\end{equation}

Under the barrier but away from the turning points, the WKB
wavefunction is given by
\begin{equation}
\begin{split}
\psi(x) = \pi^{-1/2}|k(x)|^{-1/2}
\exp(\int_{a}^{x}|k(x^{\prime})|dx^{\prime})\sin(\phi)- \\
\frac{1}{2}|k(x)|^{-1/2} \exp(-\int_{a}^{x} |k(x^{\prime})|
dx^{\prime}) \cos(\phi)
\end{split}
\end{equation}
for some value of $\phi$.

Setting $\int_{a}^{b}|k(x^{\prime}|dx^{\prime} \equiv \Gamma$, note that
\begin{equation}
\int_{a}^{x}|k(x^{\prime})|dx^{\prime}=
\Gamma-\int_{x}^{b}|k(x^{\prime}|dx^{\prime}
\end{equation}
and for x close to $b$
\begin{equation}
\int_{b}^{x}k(x^{\prime})dx^{\prime}=\frac{2}{3}C_{2}^{1/2}(x-b)^{3/2}
\end{equation}

Connecting the asymptotic forms of the Airy functions to the WKB
solution in the forbidden region gives the solution for $x \approx b$ as
\begin{equation}
\begin{split}
\psi(x)=_{x\approx b}2 C_{2}^{-1/6} \sin(\phi) e^{\Gamma}
\text{Ai}(C_{2}^{1/3}(b-x))- \\
\frac{1}{2} C_{2}^{-1/6} \cos(\phi) e^{-\Gamma} 
\text{Bi}(C_{2}^{1/3}(b-x)),
\end{split}
\end{equation}
giving a scattering phaseshift $\delta$ of
\begin{equation}
\delta = \tan^{-1}(4 e^{2 \tau} \tan(\phi))
\end{equation}
yielding a resonance at $\phi \approx 0$, $\delta \approx \frac{pi}{2}$

Similar logic gives the wavefunction for $x\approx a$
\begin{equation}
\begin{split}
\psi(x)=_{x \approx a} C_{1}^{-1/6} \sin(\phi)
\text{Bi}(C_{1}^{1/3}(x-a))- \\
C_{1}^{1/6} \cos(\phi)\text{Ai}(C_{1}^{1/3}(x-a))
\end{split}
\end{equation}

Finally, setting $\phi=0$, the ratio of the tunneling
wavefunction at the outer turning point to the tunneling wavefunction
at the inner turning point is
\begin{equation}
\frac{\psi(x=b)}{\psi(x=a)}=\frac{1}{2}e^{-\Gamma}(|\frac{C_{1}}{C_{2}}|)^{1/6}
\frac{\text{Bi}(0)}{\text{Ai}(0)}
\end{equation}


\end{document}